\documentclass[prl,twocolumn,superscriptaddress]{revtex4-2}
\usepackage{amsmath,bbm,amssymb,amsfonts,bm,color,graphicx,tabularx,physics}
\usepackage{balance}
\usepackage[unicode=true,colorlinks=true]{hyperref}
\usepackage[etex=true,export]{adjustbox}

\hypersetup{linkcolor=blue,citecolor=blue,urlcolor=blue}
\usepackage{diagbox}
\usepackage{color}

\newcommand{\sign}{{\rm sign}}

\newcommand{\beq}{\begin{equation}}
	\newcommand{\eeq}{\end{equation}}
\newcommand{\beql}{\begin{equation*}}
	\newcommand{\eeql}{\end{equation*}}
\newcommand{\beqn}{\begin{eqnarray}}
	\newcommand{\eeqn}{\end{eqnarray}}

\renewcommand{\vec}[1]{\mbox{\boldmath$#1$}}
\usepackage{comment}

\begin{document}
	
	\title{Non-Abelian line graph: A generalized approach to flat bands}
	\author{Rui-Heng Liu}
	\affiliation{School of Physics, Huazhong University of Science and Technology, Wuhan, Hubei 430074, China}
	\author{Xin Liu}
	\email{phyliuxin@hust.edu.cn}
	\affiliation{School of Physics, Huazhong University of Science and Technology, Wuhan, Hubei 430074, China}
	\affiliation{Institute for Quantum Science and Engineering and Wuhan National High Magnetic Field Center, Huazhong University of Science and Technology, Wuhan, Hubei 430074, China}
	\affiliation{Hubei Key Laboratory of Gravitation and Quantum Physics, Wuhan, Hubei 430074, China}
	
	\date{\today}
	
	\begin{abstract}
		Flat bands (FBs) in materials can enhance the correlation effects, resulting in exotic phenomena. Line graph (LG) lattices are well known for hosting FBs with isotropic hoppings in $s$-orbital models. Despite their prevalent application in the Kagome metals, there has been a lack of a general approach for incorporating higher-angular-momentum orbitals with spin-orbit couplings (SOCs) into LGs to achieve FBs. Here, we introduce a non-Abelian LG theory to construct FBs in realistic systems, which incorporates internal degrees of freedom and goes beyond $s$-orbital models. We modify the lattice edges and sites in the LG to be associated with arbitrary Hermitian matrices, referred to as the multiple LG. A fundamental aspect involves mapping the multiple LG Hamiltonian to a tight-binding (TB) model that respects the lattice symmetry through appropriate local non-Abelian transformations. We establish the general conditions to determine the local transformations. Based on this mechanism, we demonstrate the realization of $d$-orbital FBs in the Kagome lattice, which could serve as a minimal model for understanding the FBs in transition metal Kagome materials. Our approach bridges the gap between the known FBs in pure lattice models and their realization in multi-orbital systems. 
	\end{abstract}
	
	\maketitle
	
	The quenching of kinetic energy in the flat band (FB) system renders it an ideal platform for investigating diverse correlated phenomena such as ferromagnetism \cite{mielke1991ferromagnetic,mielke1991ferromagnetism}, superconductivity and superfluidity \cite{peotta2015superfluidity,xie2020topology,peri2021fragile,torma2022superconductivity}, Wigner crystal \cite{wu2007flat} and fractional quantum Hall effect \cite{sun2011nearly,neupert2011fractional,tang2011high,regnault2011fractional,andrews2020fractional,park2023observation,cai2023signatures,xu2023observation}. The destructive interference is a common origin of the FB, which suppress the kinetic energy and leads to the localization of the electrons \cite{bergman2008band,liu2014exotic}. In an early seminal work, Mielke identified a special class of lattice known as the line graph (LG) lattice that could naturally lead to destructive interference and therefore produce a FB \cite{mielke1991ferromagnetic,mielke1991ferromagnetism}, as guaranteed by the LG theorem \cite{liu2014exotic,cvetkovic2004spectral,kollar2020line}. The typical examples are Kagome \cite{mielke1992exact,bergman2008band}, checkerboard \cite{sun2011nearly,iskin2019origin} and pyrochlore lattice \cite{guo2009three,trescher2012flat,bergman2008band}. Recently, realizing FBs in real systems has attracted significant attention in both theoretical and experimental research, where transition-metal-based Kagome materials emerge as a prominent representative. Notably, both DFT calculations and experiments reveal that these materials have higher-angular-momentum orbitals (high orbitals) near the Fermi surface, along with the spin-orbit coupling (SOC) \cite{ye2018massive,kang2020dirac,liu2020orbital,yin2020fermion,kang2020topological,teng2023magnetism,arachchige2022charge,cao2023competing,jiang2023kagome,li2021observation,chen2021double,chen2022charge,hu2022rich,yang2023observation,yin2022topological,jiangkun2023kagome,cai2023energy}, which significantly differ from the isotropic hoppings of $s$-electrons as originally assumed. Meanwhile, FBs are also observed in Moiré systems, arising from the Moiré potential and band folding. Twisted bilayer graphene hosts approximate chiral FBs near charge neutrality \cite{cao2018unconventional,cao2018correlated,tarnopolsky2019origin,bernevig2021twisted1,song2021twisted,bernevig2021twisted3,lian2021twisted,bernevig2021twisted5,xie2021twisted}. More interestingly, the bands in twisted TMD systems exhibit high-orbital characteristics \cite{angeli2021gamma,xian2021realization,claassen2022ultra}, which is reminiscent of the honeycomb lattice with $p_x,p_y$ orbitals \cite{wu2007flat,wu2008p}. However, the original LG approach focuses on pure lattice degree of freedom with isotropic hoppings, and its application to systems with internal degrees of freedom (such as orbitals) remains to be studied. 
	
\begin{figure} [htbp]
		\centering
		\includegraphics[width= 1\columnwidth]{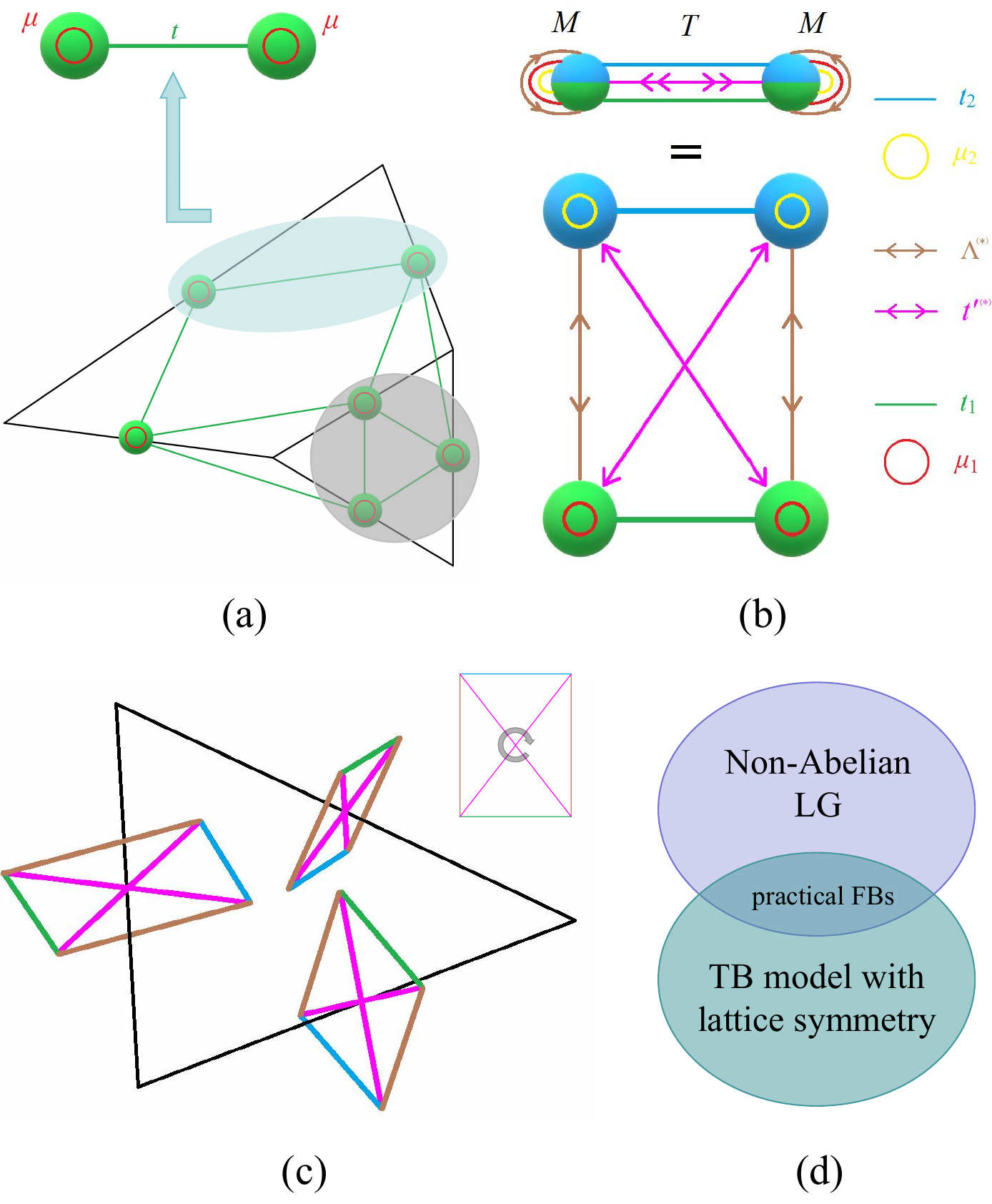}
		\caption{(a) The construction of a LG. Black and green lines belong to the root graph and the LG. The green edges and red loops represent hoppings and on-site energies in the TB model. The ``dumbbell'' is employed for convenience.
			(b) The multiple LG. The couplings are generalized to matrices.
			(c) Schematic depiction of a non-Abelian LG. The triangle corresponds to the gray region of (a). The couplings in different directions are anisotropic but related by ``rotations".
			(d) Relation between the non-Abelian LG and TB models with lattice symmetry. FBs are present in their intersection.}
		\label{intro}
	\end{figure}

	In this work, we develop a non-Abelian LG theory to incorporate internal degrees of freedom. We first generalize the uniform coupling constants (Fig.~\ref{intro}(a)) to Hermitian matrices (Fig.~\ref{intro}(b)). The extension is referred to as the multiple LG, which allows the application of the LG theorem to systems with internal space. When the LG corresponds to a periodic system, the multiple LG serves as the counterpart of the conventional LG to preserve the FB characteristic. To match the anisotropic hoppings in real materials, we perform local U$(n)$ unitary transformations in the internal space to drive the Hamiltonian beyond the isotropic multiple LG, which gives rise to a non-Abelian LG (Fig.~\ref{intro}(c)). Further, we set up general conditions to distinguish non-Abelian LGs from the tight-binding (TB) models (Fig.~\ref{intro}(d)), where all the realistic lattice symmetries are respected. To demonstrate our theory, we explicitly build and discuss $d$-orbital Kagome models, which could provide insights into Kagome materials with multi-orbital nature.
	
	\textit{Multiple Line Graph-} Mathematically, a graph consists of a set of vertices and edges, where the edge represents a pairing relation between the vertices \cite{west2001introduction} (Fig.~\ref{intro}(a)). For a simple graph $X$, we can obtain its LG $L(X)$ by replacing each edge with a vertex, which is placed on the midpoint as agreed. These vertices are adjacent when the corresponding edges in $X$ share a common vertex. $X$ is referred to as the root graph. Fig.~\ref{intro}(a) shows an example of the construction. The LG is related to a TB Hamiltonian of the form $H_{\rm LG}=tA_{L(X)}=t\sum_{i\neq j}c_{i}^{\dagger}c_{j}$. Here, $A_{L(X)}$ is the $\{0,1\}$ off-diagonal adjacency matrix of the LG and $c_i \text{ }(c^{\dagger}_i)$ the annihilation (creation) operator on the $i$-th site and $t$ the hopping constant. Since the edges are identical, there is only a unique hopping parameter $t$ between two connected vertices in the model, and it has to be real to guarantee Hermicity. When the graph is periodic, the spectrum of the LG lattice hosts a macroscopic degenerate subspace, which corresponds to a FB with energy $E_{\text{FB}}=-2t$ in the language of band theory \cite{cvetkovic2004spectral,kollar2020line}. It is commonly referred to as the LG theorem (Illustrative examples are given in the SM \cite{supp}), and the existence of FBs in the LG lattice is fundamentally governed by its specific connectivity. The uniform chemical potential $\mu$ is usually omitted in the LG since it only shifts all the bands trivially. But it is present as a loop (Fig.~\ref{intro}(a)) here for our generalization of the conventional LG to include the internal degree of freedom. The LG Hamiltonian only has two parameters $t$ and $\mu$, which are visualized as the ``dumbbell" in Fig.~\ref{intro}(a). 
	
\begin{figure} [htbp]
		\centering
		\includegraphics[width= 1\columnwidth]{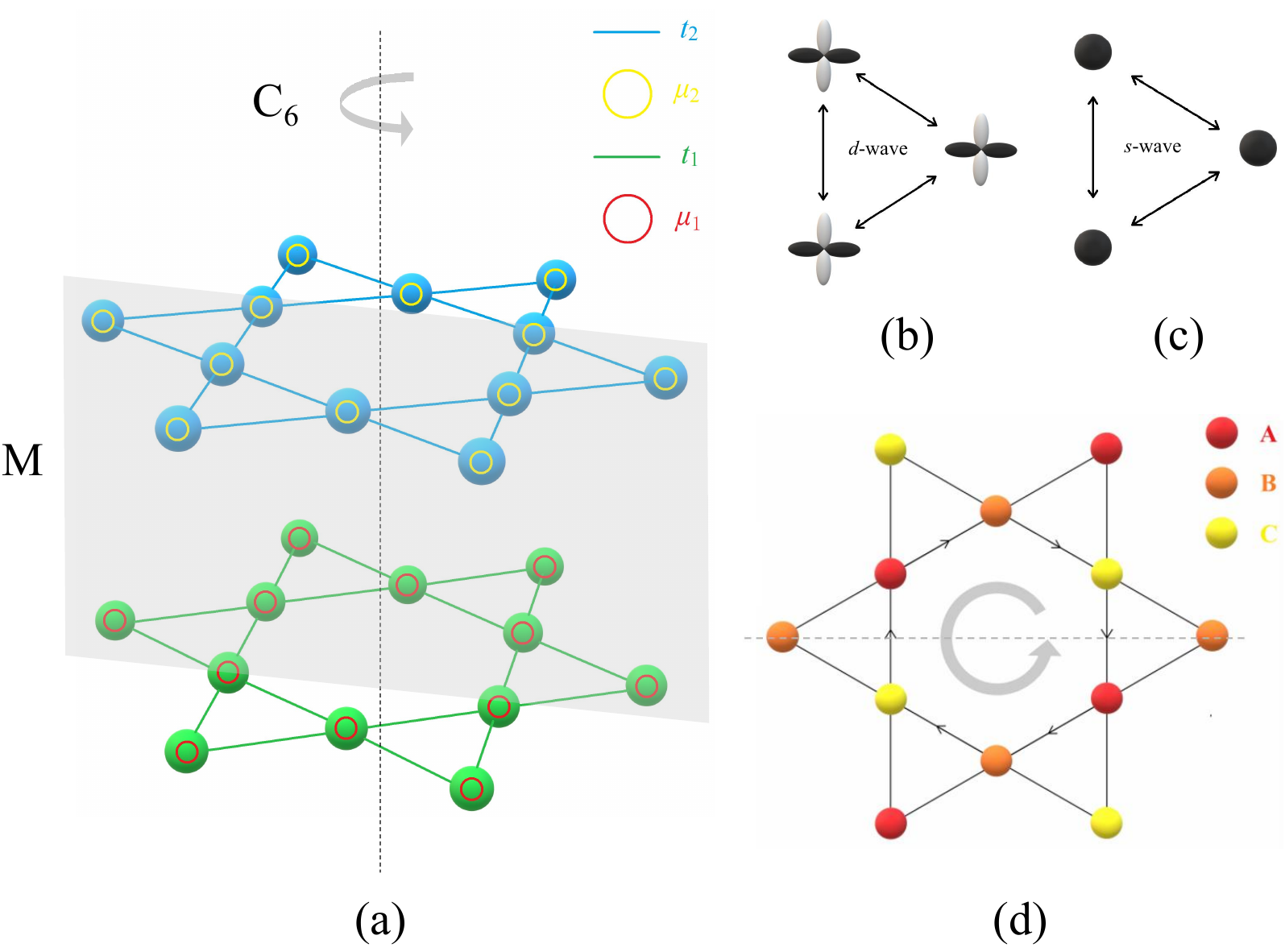}
		\caption{(a) The multiple LG on a Kagome lattice with a two-dimensional internal space, where the inter-layer coupling is neglected. The generators of the point group $C_{6v}$ are shown. 
			(b) Anisotropic hoppings from $d$-orbitals.
			(c) Isotropic hoppings from $s$-orbitals.
			(d) Sublattices A,B,C and the inequivalent nearest neighbor hoppings of Kagome.
		}
		\label{multi}
	\end{figure}

	In its essence, we extend the discussion to more general TB models with internal degrees of freedom. In such cases, the coupling between two sites is generalized to a matrix. We propose the Hamiltonian
	\begin{equation}\label{H}
		H_{\rm MLG}=T\otimes A_{L(X)}+M\otimes\mathbf{1}
	\end{equation}
	with $A_{L(X)}$ the $\{0,1\}$ off-diagonal adjacency matrix of the LG, $\mathbf{1}$ the identity of the same dimensionality with $A_{L(X)}$, $T$ and $M$ the arbitrary Hermitian matrices of dimensionality $n$. The original LG is a specific case where $T=t$ and $M=\mu$. The FB eigenspace of the original LG is spanned by the compact localized states (CLSs) and the non-contractible loop states (NLSs) \cite{bergman2008band,rhim2019classification,chen2023decoding} which are denoted as $\{v_i\}$ with $i$ up to the dimension of the FB eigenspace. To inherit the FB character from the LG, we construct the wave function $\psi_{j,i}=u_j\otimes v_i$ where $u_{j}$ is the $j$-th eigenvector, satisfying $\left(T-\dfrac{1}{2}M\right)u_{j}=\lambda_{j}u_{j}$, in the internal space. Remarkably, it follows that for arbitrary CLS or NLS $v_i$, $H\psi_{j,i}=(-2\lambda_{j})\psi_{j,i}$ holds. As a result, the original FB with energy $-2$ has transformed into $n$ FBs, each with an energy of $-2\lambda_{j}$. Their existence is independent of the parameters of the matrices $T$ and $M$, indicating that they originate from the connectivity of the underlying periodic graph. Therefore, we can naturally extend the original LG structure (Fig.~\ref{intro}(a)) to a version with multiple internal degree of freedom (Fig.~\ref{intro}(b)). Specifically, we utilize the Kagome lattice, whose root graph is the honeycomb lattice to illustrate. A single site is now mapped to multiple vertices, which results in a layer structure. We first set $T$ and $M$ to be diagonal for clearance and draw the corresponding graph structure in Fig.~\ref{multi}(a). The blue and green layers represent any two-component degrees of freedom residing on the same site. Each layer features a LG structure, while the weights could be varied in different layers as indicated by $t_{i}$ and $\mu_{i}$. By including off-diagonal elements, we can more generally incorporate inter-layer couplings (magenta and brown arrows in Fig.~\ref{intro}(b)). As long as Hermiticity is maintained, these elements can assume complex values, which could include practical factors such as the SOC or magnetic fields in principle. The distinction from the intra-layer couplings is indicated by the bidirectional arrows. It is reasonable to designate the corresponding graph structure as the multiple LG as the presence of FBs is proven.\\
	
	\textit{Physical Interpretation and Non-Abelian Line Graph-} The multiple LG provides a solid foundation for understanding FBs in LG systems with internal degrees of freedom. However, it is not directly applicable to physical high orbitals as we demonstrate below. We notice that high orbitals in the real material generally lead to an anisotropic hopping Hamiltonian (Fig.~\ref{multi}(b)), which is not compatible with the multiple LG as previously proposed. When the orbitals form a high dimensional irreducible representation (IRREP) of the symmetry group of the real crystal, there is a mismatch between symmetries and the multiple LG. Firstly, the lattice translation symmetry implies that the hoppings associated by the translation operator $\hat{T}$ are equivalent. Further, the point group imposes constraints on the inequivalent hoppings (Fig.~\ref{multi}(b)), denoted as $t^{\alpha\leftarrow\beta}$ with $\alpha,\beta$ the sublattice index. When a symmetry operation $g$ connects the hopping vectors of the inequivalent hoppings, the corresponding hopping matrices only differ by a similarity transformation as 
	\begin{equation}\label{sym_t}
		t^{\alpha\leftarrow\beta}=D^{\dagger}(g)t^{\gamma\leftarrow \delta}D(g),\quad \vec{r}^{\alpha\leftarrow\beta}=g\vec{r}^{\gamma\leftarrow \delta}
	\end{equation}
	where $D(g)$ is the representation matrix of $g$ in the high orbital basis, corresponding to a high-dimensional IRREP. In the multiple LG, the symmetry conditions are equivalent to $D^{\dagger}(g)TD(g)=T$ for $g$ is any group element. Therefore, $T$ must be proportional to the identity according to Schur's lemma. The model reduces to several identical copies of the original LG, which gives a reducible representation as the sum of the one-dimensional representations (Fig.~\ref{multi}(c)). This leads to a contradiction. Remarkably, we can remedy the loophole by utilizing the internal degree of freedom. To illustrate that, we first introduce the concept of the non-Abelian LG. The energy spectrum of the multiple LG Hamiltonian (Eq.~(\ref{H})) should be unaffected if we perform a local transformation $U_i(n)$ in the internal space. As a result, the matrix $T$ on the connecting link of the adjacent sites $i$ and $j$ changes as $T\rightarrow U_{i}^{\dagger}TU_{j}:=t^{i\leftarrow j}$ and the on-site matrix $M$ change as $M \rightarrow U^{\dagger}_i M U_i := M_i$. They are combined and understood as a gauge transformation to guarantee the invariance of the overall Hamiltonian \cite{kogut1979introduction,rothe2012lattice} and result in a Hamiltonian with enhanced local $\text{U}(n)$ symmetry
	\begin{equation}\label{tb}
		H_{\rm NALG}=\sum_{s,s'}\left(\sum_{i,j}c_{i,s}^{\dagger}t^{i\leftarrow j}_{ss'}c_{j,s'} +\sum_{i}c_{i,s}^{\dagger}M_{i}c_{i,s}\right).
	\end{equation}
	The hoppings $t^{i\leftarrow j}$  are non-Abelian matrices since the transformations $U_i(n)$ generally make them not commute to each other. Also, they are not necessarily Hermitian since each $t^{i\leftarrow j}$ only represents a one-way hopping. The corresponding graph is referred to as a non-Abelian LG (Fig.~\ref{intro}(c)). 
	
	\begin{table*}[htbp]
		\caption{The FB condition for the $d$-orbital doublet in terms of the SK integrals and the corresponding local transformation that converts the TB models to multiple LGs.}
		\begin{tabular}{ccc}
			\hline
			IRREP & FB conditions for SK integrals & Local transformation \\ \hline
			$E_1(d_{xz}/d_{yz})$ &$dd\pi+dd\delta=0$& $U_{A}=D^{2}(\text{C}_{6}),U_{B}=\mathbf{1},U_{C}=D^{-2}(\text{C}_{6})$ \\ \hline
			$E_2(d_{x^{2}-y^{2}}/d_{xy})$ &$3dd\sigma+4dd\pi+dd\delta=0$ & $U_{A}=D^{-1}(\text{C}_{6}),U_{B}=\mathbf{1},U_{C}=D(\text{C}_{6})$ \\ \hline
		\end{tabular}
		\label{local transformation}
	\end{table*}
	
\begin{figure} [htbp]
		\centering
		\includegraphics[width=1\columnwidth]{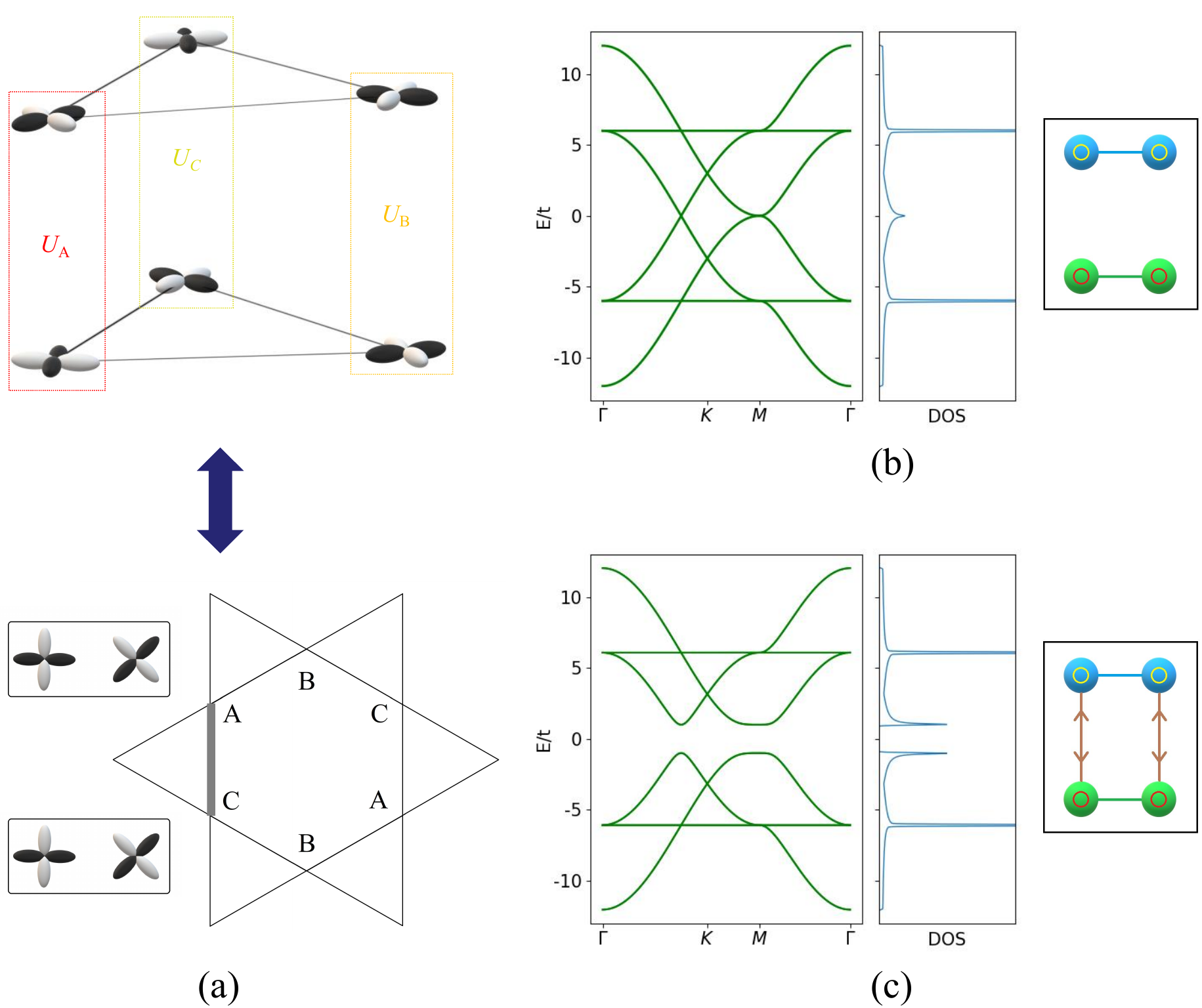}
		\caption{(a) Sublattice-dependent rotation that related the multiple LG to the $d$-orbital TB models. The grey line indicates the hopping $t^{A\leftarrow C}$.
			(b)(c) Band structure, density of states (DOS) and the corresponding multiple LG of the Kagome model with $d_{x^{2}-y^{2}}/d_{xy}$, the non-zero SK integrals are $dd\sigma=4,dd\pi=-3$.
			b) without SOC and the inter-orbital couplings; c) with SOC $\lambda=0.5$ and the inter-orbital couplings.}
		\label{rotation}
	\end{figure}

	This novel concept builds a significant connection between the multiple LG and the FB in the real material with an internal degree of freedom. A non-Abelian LG can support FBs since there always exist appropriate local transformations that can convert it back to a multiple LG. Meanwhile, the non-commutative hoppings allow it to match the high-dimensional IRREP of the space group. In practice, we can start from a TB Hamiltonian which respects all the lattice symmetry. Since the TB Hamiltonian is periodic, once we find sublattice-dependent local transformations $U_{\alpha,\beta}$ that satisfy two conditions
	\begin{equation}\label{U}
		U_{\alpha}t^{\alpha\leftarrow\beta}U_{\beta}^{\dagger}=T \quad \text{and} \quad
		U_{\alpha} M_{\alpha} U^{\dagger}_{\alpha} = M,
	\end{equation}
	the TB model is determined as a non-Abelian LG, and the presence of FBs is guaranteed. We refer to them as the hopping condition and the on-site condition, respectively. They are general regardless of the specific LG lattice or the orbitals. To better elucidate the point, we still consider the Kagome lattice for convenience. The lattice has three sublattices and six types of nearest neighbor hoppings as shown in Fig.~\ref{multi}(d). The point group of the lattice is $C_{6v}$ with the generators six-fold rotations $\text{C}_{6}$ and mirror M (Fig.~\ref{multi}(a)). Considering the smallest triangle loop in the lattice, we find that the hopping condition in Eq.~\eqref{U} leads to  
	\begin{equation}\label{tildeT}
		\tilde{T}_{\alpha}=U^{\dagger}_{\alpha}T^{3}U_{\alpha},\quad
		\tilde{T}_{\alpha}=\tilde{T}_{\alpha}^{\dagger},
	\end{equation}
	where $\tilde{T}_{\alpha=A,B,C}=\prod_{t\in \triangleright}(t^{\alpha\leftarrow\beta}\cdots t^{\gamma\leftarrow\alpha})$ is the product of the hopping matrices along the loop. All the $\tilde{T}_{\alpha}$ must be Hermitian and share the same eigenvalues, and this is a non-trivial requirement since $t^{\alpha\leftarrow\beta}$ are non-Abelian. Together with the on-site condition in Eq.~\eqref{U}, we offer a direct criterion for determining whether a given TB model is a non-Abelian LG. When the SOC is neglected, we can choose all the wavefunctions to be real and focus on a simplified spinless model. The hopping strength between atomic orbitals is usually captured by the Slater-Koster (SK) integrals \cite{slater1954simplified}, which are also real. It is natural to specify the orbitals as the $d$-orbital doublets since they carry the high-dimensional IRREPs $E_{n=1,2}$, and the corresponding representation matrices in Eq.~(\ref{sym_t}) are given by $D(\text{C}_{6})=\cos(n\theta)\sigma_0-i\sin (n\theta)\sigma_y$ and $D(M)=\sigma_{z}$. To be specific, we consider the $d_{x^{2}-y^{2}}/d_{xy}$ orbitals. Therefore, the hopping matrix along $y$ direction (Fig.~\ref{rotation}(a)) is given by
	\begin{equation}\label{tAC}
		t^{A\leftarrow C}=
		\begin{bmatrix}
			(3dd\sigma+dd\delta)/4 & 0\\
			0 & dd\pi
		\end{bmatrix}.
	\end{equation}
	The other two hoppings $t^{C\leftarrow B},t^{B\leftarrow A}$ are related to it via $\text{C}_{6}$ and $\tilde{T}_{\alpha}$ can then be calculated. The algebra of the Pauli matrices offers us a powerful tool to deal with these two-dimensional representations. Detailed calculation reveals that Eq.~(\ref{tildeT}) is satisfied when $\text{Tr}(t^{\alpha\leftarrow\beta})=0$ \cite{supp}. Equivalently, the FBs exist with the condition $3dd\sigma+4dd\pi+dd\delta=0$. Meanwhile, since $d_{x^2-y^2}/d_{xy}$ orbitals carry a two-dimensional IRREP, the on-site matrix $M$ is identity. Hence, the on-site condition in Eq.~\eqref{U} is automatically fulfilled. Notably, $t^{A\leftarrow C}\sim\sigma_{z}$ while $\sigma_{x}$ appears in the other two hoppings with these specific SK integrals. They do not commute and the model is a non-Abelian LG. Interestingly, it is close to the estimated parameters of transition metals with $|dd\sigma|>|dd\pi|>|dd\delta|$ and $\sign(dd\sigma)=-\sign(dd\pi)=\sign(dd\delta)$ \cite{masuda1984calculation}, which suggests that our proposal may be a candidate to explain the origin of FBs in realistic Kagome materials, considering the orbital components. We set $dd\sigma=4,dd\pi=-3,dd\delta=0$ and $M=0$ for simplicity and plot the band dispersion in Fig.~\ref{rotation}(b). We can manifestly identify two sets of $s$-orbital-like Kagome bands \cite{bergman2008band} in the band structure, which inherit the prominent features of the conventional LG (FB, van-Hove singularity and Dirac point). More interestingly, the van-Hove singularity and FB coexist at $M$ point in our model, which holds as long as the FB condition is satisfied.
	
	To further illustrate this relation, we explicitly transform the non-Abelian LG back to a multiple LG. It will also provide a more intuitive understanding of the band structures after we distinguish the non-Abelian LG in general TB models. Based on Eq.~\eqref{tildeT}, we further diagonalize $T$ to $\mathcal{T}=\mathcal{U}^{\dagger}T\mathcal{U}$ and obtain $\tilde{T}_{\alpha} = O^{\dagger}_{\alpha}\mathcal{T}^3 O_{\alpha}$, where $O_{\alpha}=\mathcal{U}U_{\alpha}$. Denoting the eigenvalues of $\tilde{T}_{\alpha}$ as $\text{Eig}(\tilde{T}_{\alpha})=\{\epsilon_1,\cdots, \epsilon_n\}$, it follows that $\mathcal{T}=\text{diag}(\sqrt[3]{\epsilon_1},\cdots,\sqrt[3]{\epsilon_n})$. In our $d$-orbital case, we have $\text{Eig}(\tilde{T}_{\alpha})=\{(\pm dd\pi)^{3}\}$, leading to $\mathcal{T}=\text{diag}(dd\pi,-dd\pi)$. Therefore, the related multiple LG Hamiltonian is determined as diagonal $T=\mathcal{T}=t\sigma_{z}$ for convenience, where $t=dd\pi$ and $\sigma$ acts on the orbital-space. The corresponding local transformations are listed in Table \ref{local transformation} and illustrated in Fig.~\ref{rotation}(a). The resulting multiple LG is shown in Fig.~\ref{rotation}(b), where the two layers are interpreted as the rotated orbitals in Fig.~\ref{rotation}(a). The blue, green and magenta arrows represent real-valued hoppings, and the yellow and red loops are chemical potentials (set as 0). The Hamiltonian leads to two copies of the original LG with $E_{\text{FB}}=\pm 2t$, and the band structure is symmetric due to the emergent chiral symmetry with the unitary operator $\sigma_{y}$. They are consistent with the bands shown in Fig.~\ref{rotation}(b). 
	
	On this foundation, we duplicate the model to recover a spinful model and introduce the SOC. The leading term is given by the on-site $\vec{L}\cdot\vec{S}$ as $\bra{d_{x^{2}-y^{2}},s}\vec{L}\cdot\vec{S}\ket{d_{xy},s'}=\bra{d_{x^{2}-y^{2}}}L_{z}S_{z}\ket{d_{xy}}\delta_{ss'}$, which couples the orbitals on the same site. The band structure is plotted in Fig.~\ref{rotation}(c). Correspondingly, the related multiple LG Hamiltonian is generalized to
	\begin{equation}\label{T&M}
		T=s_{0}(t\sigma_{z}),\quad
		M=s_{z}(\lambda\sigma_{y}),
	\end{equation}
	where $s$ acts on the spin-space and $\lambda$ is the SOC strength. Since $M$ commutes with the local transformations $U_{\alpha}$ (Table \ref{local transformation}), the on-site condition in Eq.~\eqref{U} is satisfied. Meanwhile, the SOC serves as the coupling between the two sets of $s$-orbital-like bands, which is visualized as the brown arrows in the multiple LG (Fig.~\ref{rotation}(c)). As a result, the degeneracy of the Dirac point and the van-Hove singularity at the charge neutrality are lifted. Remarkably, the system has a nearly flat dispersion around the original van-Hove singularity which enhances the DOS at the band edge (Fig.~\ref{rotation}(c)). In this case, the band structure can no longer be explained as superpositions of the original Kagome bands. Nevertheless, the FBs are still ensured as previously discussed. Using the anti-commutative Clifford algebra, we can solve the FB energies analytically, which are slightly shifted to $\pm\sqrt{4t^{2}+\lambda^{2}}$. Notice that spin is a good quantum number of the model, the analysis holds for each spin component. The two spin sectors share the same multiple LG structure but with opposite on-site coupling strength, resulting in a two-fold degenerate band structure. Also, the bands are symmetric due to the chiral symmetry with operator $s_{x}\sigma_{y}$. Similar analyses are also valid for the $d_{xz}/d_{yz}$ doublet, and the main results are listed in Table \ref{local transformation}. \\
	
\textit{Discussion and Conclusion-} In conclusion, we have introduced a non-Abelian LG generalized from the original LG, which is adaptable for constructing multiple FBs in practical systems with internal degrees of freedom. It also goes beyond the requirement of real-valued isotropic hoppings. Besides, our proof for the multiple LG can be extended to a plethora of pure lattice models with a FB \cite{neves2024crystal}. As long as the crystal net structure hosts a macroscopic degenerate subspace, an analogous proof still holds. In special cases such as two-dimensional internal space, the energies of multiple FBs are exactly solvable and easily tunable in the model. Considering the multi-orbital nature of Kagome materials \cite{ye2018massive,kang2020dirac,liu2020orbital,yin2020fermion,kang2020topological,teng2023magnetism,arachchige2022charge,cao2023competing,jiang2023kagome,li2021observation,chen2021double,chen2022charge,hu2022rich,yang2023observation,yin2022topological,jiangkun2023kagome,cai2023energy}, we employ the Slater-Koster formalism to explicitly show the validity of our mechanism. Our model combines the well-known characteristics of Kagome bands with high orbitals. On the other hand, the internal space mentioned in Eq.~(\ref{H}) can stand for any degree of freedom independent of the lattices, not limited to	the physical atomic orbitals in this work. It defines an equivalent class $H(T,M)$, and the models that differ only by a gauge transformation result in different non-Abelian LGs but belong to the same class and share an identical energy spectrum. These equivalent Hamiltonians can be associated with the $d$-orbital doublet via proper local transformations as we discussed, or diatomic Kagome lattice \cite{zhou2019excited,sethi2021flat,sethi2024graph}, or spin \cite{nakai2022perfect} and other more. Although they are equivalent in the gauge sense, the physical realizations are distinct. By leveraging internal degrees of freedom, our work may provide new understandings and open an alternative avenue to construct and explore practical FB systems.	
	
	\begin{acknowledgments}
		\section*{Acknowledge}
		We acknowledge useful discussions with Jin-Hua Gao and Ying-Hai Wu. We acknowledge the support by the National Natural Science Foundation of China (NSFC) (Grant No.12074133).
	\end{acknowledgments}	
	
%

	\clearpage
	\begin{appendix}
		\begin{widetext}
			\begin{center}
				\begin{Large}
					\textbf{Supplemental Materials}
				\end{Large}
			\end{center}
			
			\section{LG theorem}
			In this section, we illustrate the LG theorem with concrete examples. In graph theory, a graph $X$ consists of a set of vertices $V(X)$ and edges $E(X)$. The edges represent a pairing relation, which can be weighted or directed, between the vertices. When the pairing is unweighted and undirected, and all the edges have two distinct endpoints (no loops), the graph is called a simple graph. The corresponding adjacency matrix is off-diagonal with only $\{0,1\}$ elements, where
			\begin{equation}
				(A_{X})_{ij}=
				\begin{cases}
					1\qquad \text{$ i,j $ are connected} \\
					0\qquad \text{$ i,j $ are disconnected} 
				\end{cases}.
			\end{equation} 
			We can obtain the LG $L(X)$ from a simple graph $X$ as illustrated in the main text. The original LG theorem goes as follows:\\
			
			\textit{The spectrum of the LG exhibits an eigenspace with eigenvalue $-2$, whose multiplicity (degeneracy) is equal to the maximal number of independent even cycles.}\\
			
			The theorem is well-known in the graph theory community and the readers may refer to Ref.\cite{cvetkovic2004spectral} for more details. Here, we illustrate the theorem with the LG mentioned in the main text. Number the vertices of the LG as shown in Fig.~\ref{S1}(a), the adjacency matrix is given by 
			\begin{equation}
				A_{L(X)}=
				\begin{bmatrix}
					0 & 1 & 1 & 0 & 1 & 0\\
					1 & 0 & 0 & 1 & 0 & 0\\
					1 & 0 & 0 & 1 & 1 & 1\\
					0 & 1 & 1 & 0 & 0 & 1\\
					1 & 0 & 1 & 0 & 0 & 1\\
					0 & 0 & 1 & 1 & 1 & 0\\
				\end{bmatrix}.
			\end{equation}
			\begin{figure} [htbp]
				\centering
				\includegraphics[width= 1\columnwidth]{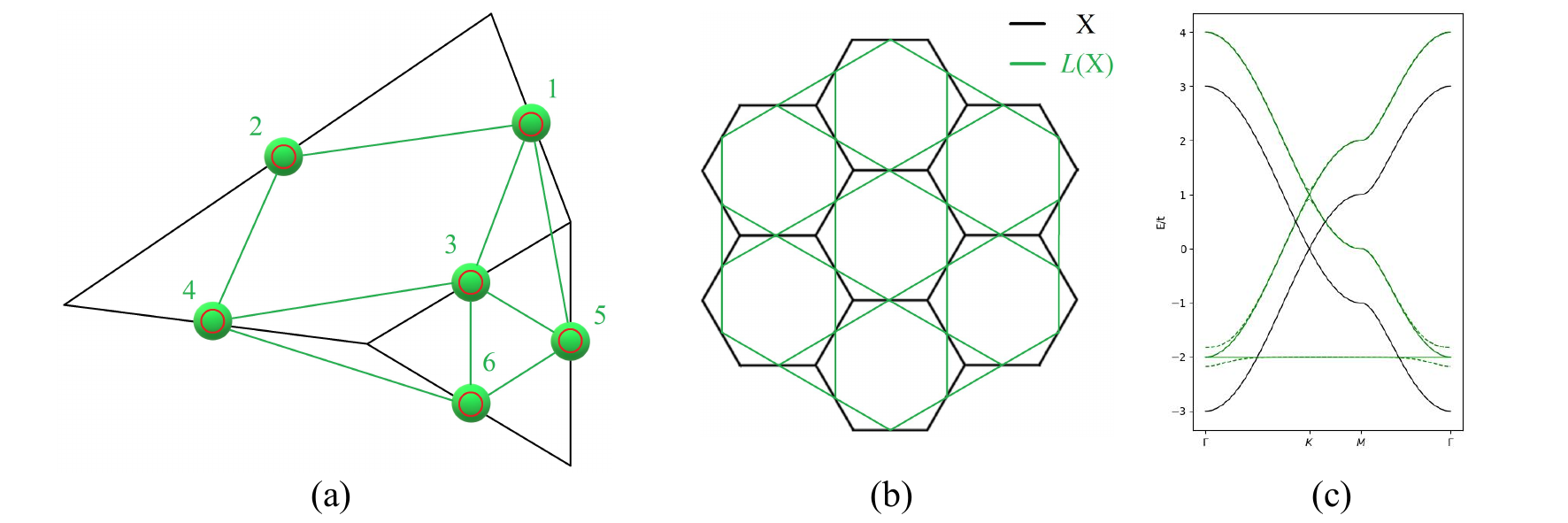}
				\caption{(a) Example of a LG, which is identical to Fig.~\ref{intro}(a). 
					(b) The honeycomb lattice and the Kagome lattice.
					(c) Band structures of the $s$-orbital honeycomb model and the Kagome model. With proper perturbations, the gaps open at $\Gamma$ and $K$ of the Kagome bands, resulting in non-trivial topology.
				}
				\label{S1}
			\end{figure}
			
			By diagonalizing the matrix numerically, the eigenvalues are
			\begin{equation}
				\text{Eig}=\{-2,-1.618,-0.861,0.618,0.746,3.115\}.
			\end{equation}
			It is explicit that the multiplicity of $-2$ is $1$. The corresponding eigenstate is 
			\begin{equation}
				[0.5,-0.5,-0.5,0.5,0,0]^{T}.
			\end{equation}
			On the other hand, the root graph has only one cycle with an even number of edges, which transforms to the vertices $1-2-4-3$ in the LG. Therefore, the LG theorem is verified in the example. When the edges are assigned an identical weight $t$, or a uniform chemical potential $\mu$ (the red loop) is included, the eigenvalues shift correspondingly and the eigenstates are intact.\\
			
			A periodic graph can also be referred to as a lattice. We can diagonalize the adjacency matrix of a LG lattice by transforming to $\vec{k}$-space, as we usually do in band theory. When the root graph $X$ is $d$-regular, which means all the vertices have a coordination number $d$, a stronger version of the LG theorem states that the spectrum of the root graph and the LG are related as 
			\begin{equation}
				E_{L(X)}=\{E_{X}+d-2,-2\}.
			\end{equation}
			The proof can be found in Ref. \cite{kollar2020line}. The eigenvalue $-2$ is macroscopic degenerate, and the related eigenstates are the CLSs and the possibly existing NLSs that span the Hilbert space of the FB. The Kagome lattice is a LG lattice whose root graph is the 3-regular honeycomb lattice (Fig.~\ref{S1}(b)). The corresponding band structures are plotted in Fig.~\ref{S1}(c).		
			
			\section{$d$-orbital Kagome model in Slater-Koster formalism}
			In this section, we give the details of the TB Hamiltonian of the $d$-orbital Kagome model. The Bravais lattice of Kagome is hexagonal with the lattice vector specified as 
			\begin{equation}
				\vec{a}_{1}=\left(\dfrac{\sqrt{3}}{2},\dfrac{1}{2}\right)\qquad
				\vec{a}_{2}=\left(0,1\right)
			\end{equation}
			where the lattice constant is set as $a=1$. The relative position of the sublattice in the unit cell is given by 
			\begin{equation}
				\vec{\tau}_{A}=0\qquad
				\vec{\tau}_{B}=\dfrac{\vec{a}_{1}}{2}\qquad
				\vec{\tau}_{C}=\dfrac{\vec{a}_{2}}{2}.
			\end{equation}
			A general TB Hamiltonian takes
			\begin{equation}
				\hat{H}=\sum_{\boldsymbol{R,R'}}\sum_{\alpha\beta}\ket{\vec{R},\alpha}t^{\alpha\beta}(\vec{R-R'})\bra{\vec{R},\beta}.
			\end{equation}
			where $\ket{\vec{R},\alpha}$ is the Wannier basis (atomic orbital) localized at $(\boldsymbol{R+\tau_{\alpha}})$, and $\alpha=1,\cdots,N_{\text{orb}}$ is the orbital label. We employ the convention that 
			\begin{equation}
				\ket{\vec{k},\alpha}=\dfrac{1}{\sqrt{N}}\sum_{\boldsymbol{R}}e^{i\boldsymbol{k}\cdot(\boldsymbol{R+\tau_{\alpha}})}\ket{\vec{R},\alpha}\qquad
				\ket{\vec{R},\alpha}=\dfrac{1}{\sqrt{N}}\sum_{\boldsymbol{R}}e^{-i\boldsymbol{k}\cdot(\boldsymbol{R+\tau_{\alpha}})}\ket{\vec{k},\alpha}
			\end{equation}
			where $\ket{\vec{k},\alpha}$ is the Bloch basis. Therefore, we can transform to $\vec{k}$-space to obtain
			\begin{equation}
				\hat{H}=\sum_{\boldsymbol{k}}\sum_{\alpha\beta}\ket{\vec{k},\alpha}H(\vec{k})_{\alpha\beta}\bra{\vec{k},\beta}
			\end{equation}
			where\begin{equation}
				H(\vec{k})_{\alpha\beta}=\sum_{\boldsymbol{R}}t^{\alpha\beta}(\vec{R})e^{-i\boldsymbol{k}\cdot(\boldsymbol{R+\tau_{\alpha}-\tau_{\beta}})}.
			\end{equation}
			The phase factor is related to the relative displacement of the orbitals. When the atomic orbitals are specified, $t^{\alpha\beta}(\vec{R})$ can be parameterized by the SK integrals.\\
			
			Considering the degenerate $d_{x^{2}-y^{2}}/d_{xy}$ doublet ($E_{2}$ IRREP of $C_{6v}$) on the Kagome lattice, the $\vec{k}$-space basis are denoted as 
			$\ket{\vec{k},d_{A,x^{2}-y^{2}}},\ket{\vec{k},d_{B,x^{2}-y^{2}}},\ket{\vec{k},d_{C,x^{2}-y^{2}}},
			\ket{\vec{k},d_{A,xy}},\ket{\vec{k},d_{B,xy}},\ket{\vec{k},d_{C,xy}}$. Limited to nearest neighbor hoppings, the non-zero matrix elements in the upper triangle are explicitly given by
			\begin{gather}
				H(\vec{k})_{12}=2\left(\dfrac{3}{16}dd\sigma+\dfrac{3}{4}dd\pi+\dfrac{1}{16}dd\delta\right)\cos\left(\dfrac{\sqrt{3}}{4}k_{x}+\dfrac{1}{4}k_{y}\right)\\
				H(\vec{k})_{13}=2\left(\dfrac{3}{4}dd\sigma+\dfrac{1}{4}dd\delta\right)\cos\left(\dfrac{1}{2}k_{y}\right)\\
				H(\vec{k})_{23}=2\left(\dfrac{3}{16}dd\sigma+\dfrac{3}{4}dd\pi+\dfrac{1}{16}dd\delta\right)\cos\left(\dfrac{\sqrt{3}}{4}k_{x}-\dfrac{1}{4}k_{y}\right)\\
				H(\vec{k})_{45}=2\left(\dfrac{9}{16}dd\sigma+\dfrac{1}{4}dd\pi+\dfrac{3}{16}dd\delta\right)\cos\left(\dfrac{\sqrt{3}}{4}k_{x}+\dfrac{1}{4}k_{y}\right)\\
				H(\vec{k})_{46}=2\left(dd\pi\right)\cos\left(\dfrac{1}{2}k_{y}\right)\\
				H(\vec{k})_{56}=2\left(\dfrac{9}{16}dd\sigma+\dfrac{1}{4}dd\pi+\dfrac{3}{16}dd\delta\right)\cos\left(\dfrac{\sqrt{3}}{4}k_{x}-\dfrac{1}{4}k_{y}\right)\\
				H(\vec{k})_{15}=H(\vec{k})_{24}=2\sqrt{3}\left(\dfrac{3}{16}dd\sigma-\dfrac{1}{4}dd\pi+\dfrac{1}{16}dd\delta\right)\cos\left(\dfrac{\sqrt{3}}{4}k_{x}+\dfrac{1}{4}k_{y}\right)\\
				H(\vec{k})_{26}=H(\vec{k})_{35}=-2\sqrt{3}\left(\dfrac{3}{16}dd\sigma-\dfrac{1}{4}dd\pi+\dfrac{1}{16}dd\delta\right)\cos\left(\dfrac{\sqrt{3}}{4}k_{x}-\dfrac{1}{4}k_{y}\right)
			\end{gather}
			When the on-site SOC is included, i.e.,
			\begin{equation}
				\bra{d_{x^{2}-y^{2}},s}\vec{L}\cdot\vec{S}\ket{d_{xy},s'}=\bra{d_{x^{2}-y^{2}}}L_{z}S_{z}\ket{d_{xy}}\delta_{ss'}.
			\end{equation}
			It corresponds to a $\sigma_{y}$ in the two-dimensional orbital space of each sublattice. Since $s_{z}$ is still a good quantum number, the spinful Hamiltonian is block-diagonal in the spin-space. The on-site coupling introduces extra terms in the $\vec{k}$-space Hamiltonian as
			\begin{equation}
				H_{\uparrow}^{\text{SOC}}(\vec{k})_{14}=
				H_{\uparrow}^{\text{SOC}}(\vec{k})_{25}=
				H_{\uparrow}^{\text{SOC}}(\vec{k})_{36}=2i\lambda.
			\end{equation}
			The Hamiltonian of the $\downarrow$ sector is obtained by $\lambda\rightarrow -\lambda$. The band structures with proper SK integrals are shown in Fig.~\ref{rotation}(b) and (c).\\
			
			Similar analyses hold for the $d_{xz}/d_{yz}$ doublet ($E_{1}$ IRREP of $C_{6v}$), where the matrix elements of the Hamiltonian are
			\allowdisplaybreaks 
			\begin{gather}
				H(\vec{k})_{12}=2\left(\dfrac{3}{4}dd\pi+\dfrac{1}{4}dd\delta\right)\cos\left(\dfrac{\sqrt{3}}{4}k_{x}+\dfrac{1}{4}k_{y}\right)\\
				H(\vec{k})_{13}=2\left(dd\delta\right)\cos\left(\dfrac{1}{2}k_{y}\right)\\
				H(\vec{k})_{23}=2\left(\dfrac{3}{4}dd\pi+\dfrac{1}{4}dd\delta\right)\cos\left(\dfrac{\sqrt{3}}{4}k_{x}+\dfrac{1}{4}k_{y}\right)\\
				H(\vec{k})_{45}=2\left(\dfrac{1}{4}dd\pi+\dfrac{3}{4}dd\delta\right)\cos\left(\dfrac{\sqrt{3}}{4}k_{x}+\dfrac{1}{4}k_{y}\right)\\
				H(\vec{k})_{46}=2\left(dd\pi\right)\cos\left(\dfrac{1}{2}k_{y}\right)\\
				H(\vec{k})_{56}=2\left(\dfrac{1}{4}dd\pi+\dfrac{3}{4}dd\delta\right)\cos\left(\dfrac{\sqrt{3}}{4}k_{x}+\dfrac{1}{4}k_{y}\right)\\
				H(\vec{k})_{15}=H(\vec{k})_{24}=2\sqrt{3}\left(\dfrac{1}{4}dd\pi-\dfrac{1}{4}dd\delta\right)\cos\left(\dfrac{\sqrt{3}}{4}k_{x}+\dfrac{1}{4}k_{y}\right)\\
				H(\vec{k})_{26}=H(\vec{k})_{35}=-2\sqrt{3}\left(\dfrac{1}{4}dd\pi-\dfrac{1}{4}dd\delta\right)\cos\left(\dfrac{\sqrt{3}}{4}k_{x}-\dfrac{1}{4}k_{y}\right)
			\end{gather}
			The band structure is analogous to the $d_{x^{2}-y^{2}}/d_{xy}$ case.
			
			\section{Finding proper local transformations}
			In this section, we show the derivation of the FB conditions and local transformations listed in Table \ref{local transformation}. For the spinless degenerate $d_{x^{2}-y^{2}}/d_{xy}$ doublet, the on-site condition in Eq.~\eqref{U} is automatically fulfilled. To satisfy Eq.~\eqref{tildeT}, we first calculate $\tilde{T}_{\alpha}$
			\begin{equation}
				\tilde{T}_{A}=t^{A\leftarrow C}t^{C\leftarrow B}t^{B\leftarrow A}\qquad
				\tilde{T}_{B}=t^{B\leftarrow A}t^{A\leftarrow C}t^{C\leftarrow B}\qquad
				\tilde{T}_{C}=t^{C\leftarrow B}t^{B\leftarrow A}t^{A\leftarrow C}.
			\end{equation}
			Invoking the symmetry constraint Eq.~\eqref{sym_t}, the hoppings to different directions are related as 
			\begin{equation}
				t^{A\leftarrow C}=D^{\dagger}(\text{C}_{6})t^{C\leftarrow B}D(\text{C}_{6})\qquad
				t^{C\leftarrow B}=D^{\dagger}(\text{C}_{6})t^{B\leftarrow A}D(\text{C}_{6})\qquad
				t^{B\leftarrow A}=D^{\dagger}(\text{C}_{6})t^{A\leftarrow C}D(\text{C}_{6})
			\end{equation}
			where\begin{equation}
				D(\text{C}_{6})=
				\begin{bmatrix}
					\cos\left(2\pi/3\right) & -\sin\left(2\pi/3\right)\\
					\sin\left(2\pi/3\right) & \cos\left(2\pi/3\right)
				\end{bmatrix}
				=\cos\left(2\pi/3\right)\sigma_0-i\sin\left(2\pi/3\right)\sigma_y
				=d_{0}\sigma_0+id_{y}\sigma_y
			\end{equation}
			Notice that $D^{3}(\text{C}_{6})=\mathbf{1}$ and $D^{\dagger}=D^{-1}$, we recognize that 
			\begin{equation}
				\tilde{T}_{C}=\left[D(\text{C}_{6})t^{A\leftarrow C}\right]^{3}
			\end{equation}
			and $\tilde{T}_{A,B}$ differ by a similarity transformation. $t^{A\leftarrow C}$ is the hopping matrix along $y$ direction which only has diagonal elements (Eq.~\eqref{tAC}). We now prove $\text{Tr}(t^{\alpha\leftarrow\beta})=0$. Generally, we can write $t^{A\leftarrow C}=t_{0}\sigma_{0}+t_{z}\sigma_{z}$ and
			\begin{equation}
				D(\text{C}_{6})t^{A\leftarrow C}
				=(d_{0}t_{0})\sigma_{0}+(d_{y}t_{z})\sigma_{x}+(id_{y}t_{0})\sigma_{y}+(d_{0}t_{z})\sigma_{z}
				=h_{0}\sigma_{0}+\vec{h}\cdot\vec{\sigma}
			\end{equation}
			Therefore\begin{equation}
				\tilde{T}_{C}=h_{0}(h_{0}^{2}+3|\vec{h}|^{2})\sigma_{0}+(3h_{0}^{2}+|\vec{h}|^{2})\vec{h}\cdot\vec{\sigma}
			\end{equation}
			is not Hermitian since $h_{y}=id_{y}t_{0}$. To meet Eq.~\eqref{tildeT}, $h_{y}$ must vanish. Since $d_{y}\neq0$, we conclude that $\text{Tr}(t^{\alpha\leftarrow\beta})=0$, which gives rise to $3dd\sigma+4dd\pi+dd\delta=0$ regarding the SK integrals. By direct calculation, the eigenvalues of $\tilde{T}_{\alpha}$ are $\text{Eig}(\tilde{T}_{\alpha})=\{(\pm dd\pi)^{3}\}$. Following the procedure in the main text, we reach at Table \ref{local transformation}. Similar analyses hold for the $d_{xz}/d_{yz}$ doublet, except that $D(\text{C}_{6})=\cos\left(\pi/3\right)\sigma_0-i\sin\left(\pi/3\right)\sigma_y,D^{3}(\text{C}_{6})=-\mathbf{1}$ and the tedious algebra is slightly different.\\
			
			It is worth noticing that we could find different local transformations to convert a non-Abelian LG back to different multiple LGs. For example, we can set 
			$U_{A}=\mathcal{U}D^{-1}(\text{C}_{6}),U_{B}=\mathcal{U},U_{C}=\mathcal{U}D(\text{C}_{6})$, where $\mathcal{U}$ is a global unitary matrix. It results in a multiple LG with non-diagonal $T$. Nevertheless, the energy spectrum is unaffected, and we could make the choice to simplify the procedure.\\
			
			After including the SOC (Eq.~\eqref{T&M}), the on-site coupling matrix $M$ can not be ignored. We can investigate two spin sectors independently and the hopping condition in Eq.~\eqref{tildeT} is satisfied as previously discussed. Since $M=s_{z}(\lambda\sigma_{y})$, $U_{\alpha}$ only involve $\sigma_{0},\sigma_{y}$, they commute with each other and the on-site condition in Eq.~\eqref{U} is satisfied. The spinful TB model is still a non-Abelian LG, with $s_{0}U_{\alpha}$ converting it back to a multiple LG.
			
		\end{widetext}
	\end{appendix}

\end{document}